\documentclass[final]{ws-ijgmmp}%
\usepackage{amsmath}
\usepackage{amsbsy}
\usepackage{amsfonts}
\usepackage{cite}
\usepackage{amssymb}%
\numberwithin{equation}{section}
\def\theequation{\arabic{section}.\arabic{equation}}
\normalfont\upshape
\begin{document}
\markboth{Steven Duplij}{Formulation of singular theories
in a partial Hamiltonian formalism}
%
\catchline{}{}{}{}{}
%

\author{{Steven Duplij}}

\address{Mathematisches Institut,
Universit\"at M\"unster\\
Einsteinstr. 62,
D-48149 M\"unster, Germany\\
{duplijs@math.uni-muenster.de}}

\title{\textbf{Formulation of singular theories in a partial Hamiltonian formalism
using a new bracket and multi-time dynamics}}
\maketitle

\begin{history}
\received{(6 August, 2013)}
\accepted{(20 August 2014)}
Published
\end{history}

\begin{abstract}
A formulation of singular classical theories (determined by degenerate
Lagrangians) without constraints is presented. A partial Hamiltonian formalism in
the phase space having an initially arbitrary number of momenta (which can be
smaller than the number of velocities) is proposed. The equations of motion become
first-order differential equations, and they coincide with those of multi-time
dynamics, if a certain condition is imposed. In a singular theory, this
condition is fulfilled in the case of the coincidence of the number of
generalized momenta with the rank of the Hessian matrix. The noncanonical
generalized velocities satisfy a system of linear algebraic equations, which
allows an appropriate classification of singular theories (gauge and
nongauge). A new antisymmetric bracket (similar to the Poisson bracket) is
introduced, which describes the time evolution of physical quantities in a
singular theory. The origin of constraints is shown to be a consequence of the
(unneeded in our formulation) extension of the phase space, when the
new bracket transforms into the Dirac bracket. Quantization is briefly discussed.

\end{abstract}


\begin{small}

\tableofcontents
\end{small}

\section{Introduction}

Many modern physical models are gauge theories (see, for example,
\cite{weinberg123}), which are described at the classical level by singular
(degenerate) Lagrangians \cite{car6}. The transition to the normally used
sequential quantum Hamiltonian formalism for such singular theories is
non-trivial (because it is not possible to directly apply the Legendre
transformation \cite{tul, tul/urb}), and requires additional constructions
\cite{men/tul, mar/men/tul}. The main difficulty lies in the appearance of
additional relations between the dynamical variables, which are called
constraints \cite{dirac}. These are used to construct a reduced phase space
with fewer positions and corresponding momenta. Further, the selection of the
physical subspace of the reduced phase space), where one can consistently
carry out the procedure of quantization is needed. \cite{git/tyu, hen/tei}.
Despite the widespread use of constraint theory \cite{sundermeyer, reg/tei},
it is not free from internal contradictions and problems \cite{pon3, mik/zan}.
So it makes sense to revise the Hamiltonian formalism itself for singular
theories with degenerate Lagrangians \cite{dup2009, dup2011}.

The purpose of this paper is to describe singular theories without the help of
constraints. Firstly, a partial Hamiltonian formalism for any Lagrangian
system is constructed with an arbitrary number of momenta which can be less
than the number of velocities. The corresponding system of equations of motion
derived from the principle of least action contains the first derivatives of
the canonical variables and second-order derivatives of the generalized
noncanonical coordinates (for their velocities we do not define corresponding
momenta). Under certain conditions, the equations for the latter are
differential-algebraic equations of first order; such a physical system is
equivalent to a multi-time dynamics (see Appendix). These conditions exist in
theories with degenerate Lagrangians, if the number of momenta is equal to the
rank of the Hessian. Then the equations for the noncanonical generalized
velocities are algebraic, and the dynamics is defined in terms of new
bracket, which, like the Poisson bracket, is antisymmetric and satisfies the
Jacobi identity. In this formalism, there are no additional relations between
the dynamical variables (constraints). It is shown that, if we extend the
phase space so that additional (\textquotedblleft extra\textquotedblright)
momenta corresponding to the noncanonical generalized velocities are defined,
the constraints would appear, a new bracket turns into the Dirac bracket, 
and suitable formulas reproduce the Dirac theory
\cite{dirac}. For clarity, we use local coordinates and consider a system with
a finite number of degrees of freedom.

\section{Preliminaries}

A dynamical system (with a finite number of degrees of freedom) can be defined
in terms of generalized (Lagrangian) coordinates $q^{A}\left(  t\right)  $,
$A=1,\ldots,n$ (as a function of time $t$) in the configuration space $Q_{n}$
(we write its dimension $n$ as a lower index). The trajectory in the
configuration space $Q_{n}$, is determined by the equations of motion, which
is a set of differential equations for the generalized coordinates
$q^{A}\left(  t\right)  $, and their time derivatives $\dot{q}^{A}\left(
t\right)  $, where $\dot{q}^{A}\left(  t\right)  \equiv dq^{A}\left(
t\right)  \diagup dt$, which define the tangent bundle $TQ_{n}$ of rank $n$
(so that the dimension of the total space is $2n$) \cite{arnold}. Here we do
not consider systems with higher derivatives (see, e.g.,
\cite{nak/ham,and/gon/mac/mas}). The equations of motion can be obtained using
different principles of action, which identify the actual trajectory with the
requirement of a functional extremity \cite{lanczos}. The standard principle
of least action \cite{lanczos} considers the functional
\begin{equation}
S=\int_{t_{0}}^{t}Ldt^{\prime},\label{sl}%
\end{equation}
where a differentiable function $L=L\left(  t,q^{A},\dot{q}^{A}\right)  $ is a
Lagrangian, and the functional (on extremals) $S=S\left(  t,q^{A}\right)  $ is
the action of a dynamical system as a function of the upper limit of $t$ (a
fixed lower limit $t_{0}$). Let us consider an infinitesimal variation of the
functional (\ref{sl}) $\delta S=S\left(  t+\delta t,q^{A}+\delta q^{A}\right)
-S\left(  t,q^{A}\right)  $. Without loss of generality, we can assume that
the lower limit of the variation vanishes $\delta q^{A}\left(  t_{0}\right)
=0$, while the upper denotes the trajectory $\delta q^{A}$. For the variation
of $\delta S$ after integration by parts \footnote{Repeated upper and lower
indices imply summation. Indices in the function arguments are not to be
summed, rather they are written explicitly to distinguish between different
types of variables.}
\begin{equation}
\delta S=\int_{t_{0}}^{t}\left(  \dfrac{\partial L}{\partial q^{A}}-\dfrac
{d}{dt^{\prime}}\left(  \dfrac{\partial L}{\partial\dot{q}^{A}}\right)
\right)  \delta q^{A}\left(  t^{\prime}\right)  dt^{\prime}+\dfrac{\partial
L}{\partial\dot{q}^{A}}\delta q^{A}+\left(  L-\dfrac{\partial L}{\partial
\dot{q}^{A}}\dot{q}^{A}\right)  \delta t,\label{ds}%
\end{equation}
Then, from the principle of least action of $\delta S=0$, we obtain the
equations of motion (Euler-Lagrange equations) \cite{lan/lif1}
\begin{equation}
\dfrac{\partial L}{\partial q^{A}}-\dfrac{d}{dt}\left(  \dfrac{\partial
L}{\partial\dot{q}^{A}}\right)  =0\ \ \ A=1,\ldots,n,\label{le}%
\end{equation}
which determine the extremals under the condition that all $\delta q^{A}=0$
and $\delta t=0$. The second and third terms in (\ref{ds}) define the total
differential of the action (on extremals) as a function of $\left(
n+1\right)  $ variables: the coordinates and the upper limit of integration in
(\ref{sl})
\begin{equation}
dS=\dfrac{\partial L}{\partial\dot{q}^{A}}dq^{A}+\left(  L-\dfrac{\partial
L}{\partial\dot{q}^{A}}\dot{q}^{A}\right)  dt.\label{ds1}%
\end{equation}
Thus, from the definition of the action (\ref{sl}) and (\ref{ds1}), it follows
that $\dfrac{dS}{dt}=L$,\ \ \ $\dfrac{\partial S}{\partial q^{A}}%
=\dfrac{\partial L}{\partial\dot{q}^{A}}$,\ \ \ $\dfrac{\partial S}{\partial
t}=L-\dfrac{\partial L}{\partial\dot{q}^{A}}\dot{q}^{A}$. In the standard
Hamiltonian formalism \cite{lan/lif1}, to each coordinate $q^{A}$ one can
assign the canonically conjugate momentum $p_{A}$ by the formula%
\begin{equation}
p_{A}=\dfrac{\partial L}{\partial\dot{q}^{A}},\ \ \ A=1,\ldots,n.\label{p}%
\end{equation}
If the system of equations (\ref{p}) is solvable for all velocities $\dot
{q}^{A}$, we can define the Hamiltonian by the Legendre transform
\cite{lan/lif1}
\begin{equation}
H=p_{A}\dot{q}^{A}-L,\label{hp}%
\end{equation}
which defines the mapping between the cotangent and tangent bundles
$TQ_{n}^{\ast}\rightarrow TQ_{n}$ \cite{arnold}. The r.h.s. of (\ref{hp}) is
expressed in terms of the momenta, so that $H=H\left(  t,q^{A},p_{A}\right)  $
is a function on the phase space (or the cotangent bundle $TQ_{n}^{\ast}$ of
rank $n$ and the dimension of the total space $2n$), that is independent of
$2n$ canonical coordinates $\left(  q^{A},p_{A}\right)  $. In the standard
canonical formalism, each coordinate $q^{A}$ has its conjugate momentum
$p_{A}$ by the formula (\ref{p}), we call it a full Hamiltonian formalism (in
the full phase space $TQ_{n}^{\ast}$). Then the differential of the action
(\ref{ds1}) can also be written as%
\begin{equation}
dS=p_{A}dq^{A}-Hdt.\label{sp}%
\end{equation}
Therefore, for partial derivatives we get
\begin{equation}
\dfrac{\partial S}{\partial q^{A}}=p_{A},\ \ \ \dfrac{\partial S}{\partial
t}=-H\left(  t,q^{A},p_{A}\right)  ,\label{spd}%
\end{equation}
which implies the Hamilton-Jacobi differential equation%
\begin{equation}
\dfrac{\partial S}{\partial t}+H\left(  t,q^{A},\dfrac{\partial S}{\partial
q^{A}}\right)  =0,
\end{equation}
which fully determines the dynamics in terms of the given Hamiltonian
$H\left(  t,q^{A},p_{A}\right)  $. Variation of the action
\begin{equation}
S=\int\left(  p_{A}dq^{A}-Hdt\right)  \label{spi}%
\end{equation}
while considering the coordinates and momenta as independent variables, and
further integration by parts leads in a standard way \cite{lan/lif1} to
Hamilton's equations in the differential form\footnote{In fact, the equations
(\ref{qp}) are the conditions for a closed differential 1-form (\ref{sp})
(Poincare-Cartan) \cite{arnold}.}
\begin{equation}
dq^{A}=\dfrac{\partial H}{\partial p_{A}}dt,\ \ \ \ \ dp_{A}=-\dfrac{\partial
H}{\partial q^{A}}dt\label{qp}%
\end{equation}
for the full Hamiltonian formalism (that is, the dynamical system is fully
specified by $TQ_{n}^{\ast}$). If we define the (full) Poisson bracket by%
\begin{equation}
\left\{  A,B\right\}  _{full}=\dfrac{\partial A}{\partial q^{A}}%
\dfrac{\partial B}{\partial p_{A}}-\dfrac{\partial B}{\partial q^{A}}%
\dfrac{\partial A}{\partial p_{A}},\label{abf}%
\end{equation}
then the equation (\ref{qp}) can be written in the standard form
\cite{lan/lif1}
\begin{equation}
dq^{A}=\left\{  q^{A},H\right\}  _{full}\ dt,\ \ \ \ dp_{A}=\left\{
p_{A},H\right\}  _{full}\ dt.\label{qpp}%
\end{equation}
It is clear that the two formulations of the principle of the least action,
that is (\ref{sl}) and (\ref{spi}), are completely equivalent (they describe
the same dynamics) from the definitions of the momenta (\ref{p}) and the
Hamiltonian (\ref{hp}). Now we will show in this simple language, how to
describe the same dynamics using fewer generalized momenta than the number of
generalized coordinates.

\section{Partial Hamiltonian formalism}

The transition from a full to a partial Hamiltonian formalism and a multi-time
dynamics can be analyzed using the following well-known classical analogy
\cite{lanczos}, but in its reverse form. In the study of the parametric form
of the canonical equations and action (\ref{spi}), one formally introduces the
extended phase space with the following additional position and momentum
\begin{align}
q^{n+1}  &  =t,\label{qn1}\\
p_{n+1}  &  =-H. \label{pn1}%
\end{align}
Then the action (\ref{spi}) takes the symmetrical form and contains only the
first term \cite{lanczos} (see, also \cite{arn/des/mis})%
\begin{equation}
S=\int\sum_{A=1}^{n+1}p_{A}dq^{A}.
\end{equation}
Here we proceed in the opposite way, and ask: can we, on the contrary, reduce
the number of momenta that describe the dynamical system, that is, can we
formulate a partial Hamiltonian formalism, which would be equivalent (at the
classical level) to the Lagrangian formalism? In other words, is it possible
to describe the system with the initial action (\ref{sl}) in a smaller phase
space initially which is not reduced, because the full phase space is not
defined at all. Can we build an analog of the action (\ref{spi}) and obtain
equations of motion which are equivalent to the Lagrange equations of motion
(\ref{le}), and what additional conditions are needed for this? It turns out
that the answer to all these questions is positive and leads to a description
of singular theories (with degenerate Lagrangians) without introducing
constraints \cite{dup2009, dup2011}.

We define a partial Hamiltonian formalism \cite{dup2013}, so that the conjugate momentum is
associated not to every $q^{A}$ by the formula (\ref{p}), but only for the
first $n_{p}<n$ generalized coordinates, which are called canonical and
denoted by $q^{i}$, $i=1,\ldots n_{p}$. The resulting manifold $TQ_{n_{p}%
}^{\ast}$ is defined by $2n_{p}$ (reduced) canonical coordinates $\left(
q^{i},p_{i}\right)  $. The rest of the generalized coordinates are called
noncanonical $q^{\alpha}$, $\alpha=n_{p}+1,\ldots n$, and they form a
configuration subspace $Q_{n-n_{p}}$, which corresponds to the tangent bundle
$TQ_{n-n_{p}}$ (a subscript indicates the corresponding dimension of the total
space). Thus, the dynamical system is now given on the direct product (of
manifolds) $TQ_{n_{p}}^{\ast}\times TQ_{n-n_{p}}$. For reduced generalized
momenta we have
\begin{equation}
p_{i}=\dfrac{\partial L}{\partial\dot{q}^{i}},\ \ \ i=1,\ldots,n_{p}.
\label{pp}%
\end{equation}
A partial Hamiltonian, similar to (\ref{hp}), is defined by a partial Legendre
transform
\begin{equation}
H_{0}=p_{i}\dot{q}^{i}+\dfrac{\partial L}{\partial\dot{q}^{\alpha}}\dot
{q}^{\alpha}-L, \label{hl0}%
\end{equation}
which defines a mapping of $TQ_{n_{p}}^{\ast}\times TQ_{n-n_{p}}\rightarrow
TQ_{n}$ (see ( \ref{hp})). In (\ref{hl0}) the canonical generalized velocities
$\dot{q}^{i}$ are expressed in terms of the reduced canonical momenta $p_{i}$
by (\ref{pp}). For the action differential (\ref{sp}) we can write%
\begin{equation}
dS=p_{i}dq^{i}+\dfrac{\partial L}{\partial\dot{q}^{\alpha}}dq^{\alpha}%
-H_{0}dt. \label{dss}%
\end{equation}
We define
\begin{equation}
H_{\alpha}=-\dfrac{\partial L}{\partial\dot{q}^{\alpha}},\ \ \ \alpha
=n_{p}+1,\ldots n \label{ha}%
\end{equation}
and call the functions $H_{\alpha}$ additional Hamiltonians, then
\begin{equation}
dS=p_{i}dq^{i}-H_{\alpha}dq^{\alpha}-H_{0}dt. \label{spq}%
\end{equation}
Note that without the second term in (\ref{spq}) the partial Hamiltonian
(\ref{hl0}) is the Routh function $R=p_{i}\dot{q}^{i}-L$, in terms of which
the Lagrange equations of motion can be reformulated \cite{lan/lif1}. However,
a consistent formulation of the principle of least action for $S$ and a
multi-time dynamics of singular systems \cite{dup2009} (see Appendix) is
natural in terms of the introduced additional Hamiltonians $H_{\alpha}$
(\ref{ha}), while coordinates $q^{\alpha}$ can be treated as additional times
(effectively, which can be observed from (\ref{spq})).

Thus, in the partial Hamiltonian formalism the dynamics is completely
determined by not one Hamiltonian $H_{0}$ only, but by the set of $\left(
n-n_{p}+1\right)  $ Hamiltonians $H_{0}$, $H_{\alpha}$, $\alpha=n_{p}+1,\ldots
n$. Indeed, it follows from (\ref{spq}) that the partial derivatives of the
action $S=S\left(  t,q^{i},q^{\alpha}\right)  $ are (see (\ref{spd}) for the
standard case)
\begin{align}
\dfrac{\partial S}{\partial q^{i}}  &  =p_{i},\\
\dfrac{\partial S}{\partial q^{\alpha}}  &  =-H_{\alpha}\left(  t,q^{i}%
,p_{i},q^{\alpha},\dot{q}^{\alpha}\right)  ,\\
\dfrac{\partial S}{\partial t}  &  =-H_{0}\left(  t,q^{i},p_{i},q^{\alpha
},\dot{q}^{\alpha}\right)  ,
\end{align}
which yields the system $\left(  n-n_{p}+1\right)  $ of Hamilton-Jacobi
equations%
\begin{align}
\dfrac{\partial S}{\partial t}+H_{0}\left(  t,q^{i},\dfrac{\partial
S}{\partial q^{i}},q^{\alpha},\dot{q}^{\alpha}\right)   &  =0,\label{st}\\
\dfrac{\partial S}{\partial q^{\alpha}}+H_{\alpha}\left(  t,q^{i}%
,\dfrac{\partial S}{\partial q^{i}},q^{\alpha},\dot{q}^{\alpha}\right)   &
=0. \label{sq}%
\end{align}
Now, on the direct product of $TQ_{n_{p}}^{\ast}\times TQ_{n-n_{p}}$, the
action is
\begin{equation}
S=\int\left(  p_{i}dq^{i}-H_{\alpha}dq^{\alpha}-H_{0}dt\right)  . \label{spq1}%
\end{equation}
Variation of (\ref{spq1}) should be made independently on $2n_{p}$-reduced
canonical coordinates $q^{i},p_{i}$ and $\left(  n-n_{p}\right)  $
noncanonical generalized coordinates $q^{\alpha}$. Under the assumption that
the variations of $\delta q^{i}$, $\delta p_{i}$, $\delta q^{\alpha}$ at the
upper and lower limits vanish after integration by parts for the variation of
the action (\ref{spq1}) we obtain
\begin{align}
\delta S  &  =\int\delta p_{i}\left[  dq^{i}-\dfrac{\partial H_{0}}{\partial
p_{i}}dt-\dfrac{\partial H_{\beta}}{\partial p_{i}}dq^{\beta}\right]
+\int\delta q^{i}\left[  -dp_{i}-\dfrac{\partial H_{0}}{\partial q^{i}%
}dt-\dfrac{\partial H_{\beta}}{\partial q^{i}}dq^{\beta}\right]  +\nonumber\\
&  \int\delta q^{\alpha}\left[  \dfrac{\partial H_{\alpha}}{\partial\dot
{q}^{\beta}}d\dot{q}^{\beta}+\dfrac{\partial H_{\alpha}}{\partial q^{i}}%
dq^{i}+\dfrac{\partial H_{\alpha}}{\partial p_{i}}dp_{i}+\dfrac{d}{dt}\left(
\dfrac{\partial H_{0}}{\partial\dot{q}^{\alpha}}+\dfrac{\partial H_{\beta}%
}{\partial\dot{q}^{\alpha}}\dot{q}^{\beta}\right)  dt+\right. \nonumber\\
&  \left.  +\left(  \dfrac{\partial H_{\alpha}}{\partial q^{\beta}}%
-\dfrac{\partial H_{\beta}}{\partial q^{\alpha}}\right)  dq^{\beta}+\left(
\dfrac{\partial H_{\alpha}}{\partial t}-\dfrac{\partial H_{0}}{\partial
q^{\alpha}}\right)  dt\right]  . \label{dst}%
\end{align}
The equations of motion for the partial Hamiltonian formalism can be derived
from the principle of the least action $\delta S=0$. Taking into account the
fact that the variations of $\delta q^{i}$, $\delta p_{i}$, $\delta q^{\alpha
}$ are independent, their coefficients (each bracket in (\ref{dst})) vanish.
We introduce the (reduced) Poisson bracket for two functions $A$ and $B$ in
the reduced phase space $TQ_{n_{p}}^{\ast}$%
\begin{equation}
\left\{  A,B\right\}  =\dfrac{\partial A}{\partial q^{i}}\dfrac{\partial
B}{\partial p_{i}}-\dfrac{\partial B}{\partial q^{i}}\dfrac{\partial
A}{\partial p_{i}}. \label{ab}%
\end{equation}
Then, substituting $dq^{i}$ and $dp_{i}$ from the first row of (\ref{dst}) in
the second line, we obtain the equations of motion on $TQ_{n_{p}}^{\ast}\times
TQ_{n-n_{p}}$ in the differential form
\begin{align}
dq^{i}  &  =\left\{  q^{i},H_{0}\right\}  dt+\left\{  q^{i},H_{\beta}\right\}
dq^{\beta},\label{dq}\\
dp_{i}  &  =\left\{  p_{i},H_{0}\right\}  dt+\left\{  p_{i},H_{\beta}\right\}
dq^{\beta},\label{dp}\\
\dfrac{\partial H_{\alpha}}{\partial\dot{q}^{\beta}}d\dot{q}^{\beta}+\dfrac
{d}{dt}\left(  \dfrac{\partial H_{0}}{\partial\dot{q}^{\alpha}}+\dfrac
{\partial H_{\beta}}{\partial\dot{q}^{\alpha}}\dot{q}^{\beta}\right)  dt  &
=\left(  \dfrac{\partial H_{\beta}}{\partial q^{\alpha}}-\dfrac{\partial
H_{\alpha}}{\partial q^{\beta}}+\left\{  H_{\beta},H_{\alpha}\right\}
\right)  dq^{\beta}\nonumber\\
&  +\left(  \dfrac{\partial H_{0}}{\partial q^{\alpha}}-\dfrac{\partial
H_{\alpha}}{\partial t}+\left\{  H_{0},H_{\alpha}\right\}  \right)  dt.
\label{dhq}%
\end{align}

We see that on $TQ_{n_{p}}^{\ast}$, we have the first-order equations
(\ref{dq})--(\ref{dp}) for the canonical coordinates $q^{i}$, $p_{i}$, as it
should be (see (\ref{qp})), while on the (noncanonical) subspace $TQ_{n-n_{p}%
}$, the equation (\ref{dhq}) is still of second order with respect to the
noncanonical generalized coordinates $q^{\alpha}$, namely,
\begin{align}
\dot{q}^{i}  &  =\left\{  q^{i},H_{0}\right\}  +\left\{  q^{i},H_{\beta
}\right\}  \dot{q}^{\beta},\label{dq1}\\
\dot{p}_{i}  &  =\left\{  p_{i},H_{0}\right\}  +\left\{  p_{i},H_{\beta
}\right\}  \dot{q}^{\beta},\label{dp1}\\
\dfrac{\partial H_{\alpha}}{\partial\dot{q}^{\beta}}\ddot{q}^{\beta}+\dfrac
{d}{dt}\left(  \dfrac{\partial H_{0}}{\partial\dot{q}^{\alpha}}+\dfrac
{\partial H_{\beta}}{\partial\dot{q}^{\alpha}}\dot{q}^{\beta}\right)   &
=\left(  \dfrac{\partial H_{\beta}}{\partial q^{\alpha}}-\dfrac{\partial
H_{\alpha}}{\partial q^{\beta}}+\left\{  H_{\beta},H_{\alpha}\right\}
\right)  \dot{q}^{\beta}\nonumber\\
&  +\left(  \dfrac{\partial H_{0}}{\partial q^{\alpha}}-\dfrac{\partial
H_{\alpha}}{\partial t}+\left\{  H_{0},H_{\alpha}\right\}  \right)  .
\label{dhq1}%
\end{align}
It is important to note that the resulting system of equations of motion
(\ref{dq1} )--(\ref{dhq1}) of the partial Hamiltonian formalism is valid for
any number of reduced momenta (which becomes a free parameter together with
the number $\left(  n-n_{p}\right)  $ of the additional Hamiltonians
$H_{\alpha}$)%
\begin{equation}
0\leq n_{p}\leq n.
\end{equation}
In other words, the dynamics is independent of the dimension of the reduced
phase space. In this case, the boundary values of $n_{p}$ correspond to the
Lagrangian and Hamiltonian formalisms, respectively, so that we have the three
cases, which are described by the equations (\ref{dq1})--(\ref{dhq1}):

\begin{enumerate}
\item $n_{p}=0$ --- the Lagrangian formalism on $TQ_{n}$ (we have only the
last equation (\ref{dhq1}) without the Poisson brackets), and $\alpha
=1,\ldots,n$;

\item $0<n_{p}<n$ --- the partial Hamiltonian formalism for $TQ_{n_{p}}^{\ast
}\times TQ_{n-n_{p}}$ (all of the equations are considered);

\item $n_{p}=n$ --- the standard Hamiltonian formalism on $TQ_{n}^{\ast}$ (the
first two equations (\ref{dq1} )--(\ref{dp1}) without the second terms
containing noncanonical generalized velocities are considered), which
coincides with the standard Hamiltonian equations (\ref{qp}), and
$i=1,\ldots,n$.
\end{enumerate}

Let us show that in the case 1) we obtain the standard Lagrange equations (for
the noncanonical variables $q^{\alpha}$). Indeed, the equation (\ref{dhq1})
without the Poisson brackets (the canonical variables $q^{i},p_{i}$ are absent
in the case $n_{p}=0$ ) can be rewritten as
\begin{equation}
\dfrac{\partial H_{\alpha}}{\partial\dot{q}^{\beta}}\ddot{q}^{\beta}+\dfrac
{d}{dt}\dfrac{\partial}{\partial\dot{q}^{\alpha}}\left(  H_{0}+H_{\beta}%
\dot{q}^{\beta}\right)  -\dfrac{dH_{\alpha}}{dt}=\dfrac{\partial}{\partial
q^{\alpha}}\left(  H_{0}+H_{\beta}\dot{q}^{\beta}\right)  -\dfrac{\partial
H_{\alpha}}{\partial q^{\beta}}\dot{q}^{\beta}-\dfrac{\partial H_{\alpha}%
}{\partial t}, \label{h2}%
\end{equation}
where we derive
\begin{align}
\dfrac{d}{dt}\left(  \dfrac{\partial H_{0}}{\partial\dot{q}^{\alpha}}%
+\dfrac{\partial H_{\beta}}{\partial\dot{q}^{\alpha}}\dot{q}^{\beta}\right)
&  =\dfrac{d}{dt}\left[  \dfrac{\partial}{\partial\dot{q}^{\alpha}}\left(
H_{0}+H_{\beta}\dot{q}^{\beta}\right)  -H_{\beta}\dfrac{\partial}{\partial
\dot{q}^{\alpha}}\dot{q}^{\beta}\right] \nonumber\\
&  =\dfrac{d}{dt}\left[  \dfrac{\partial}{\partial\dot{q}^{\alpha}}\left(
H_{0}+H_{\beta}\dot{q}^{\beta}\right)  -H_{\alpha}\right]  .
\end{align}
Using the expression for the total derivative of $dH_{\alpha}\diagup dt$, we
obtain from (\ref{h2})%
\begin{equation}
\dfrac{d}{dt}\dfrac{\partial}{\partial\dot{q}^{\alpha}}\left(  H_{0}+H_{\beta
}\dot{q}^{\beta}\right)  =\dfrac{\partial}{\partial q^{\alpha}}\left(
H_{0}+H_{\beta}\dot{q}^{\beta}\right)  . \label{ddh}%
\end{equation}
The formula to determine the partial Hamiltonian (\ref{hl0}) without variables
$q^{i},p_{i}$, taking into account (\ref{ha}) is
\begin{equation}
H_{0}=-H_{\alpha}\dot{q}^{\alpha}-L.
\end{equation}
Hence, $H_{0}+H_{\beta}\dot{q}^{\beta}=-L$, so that from (\ref{ddh}), we
obtain the standard Lagrange equations in noncanonical sector
\begin{equation}
\dfrac{d}{dt}\dfrac{\partial}{\partial\dot{q}^{\alpha}}L=\dfrac{\partial
}{\partial q^{\alpha}}L.
\end{equation}
As in the standard Hamiltonian formalism \cite{lan/lif1}, a non-trivial
dynamics in the noncanonical sector is determined by the presence of
noncanonical terms with second derivatives, that is, the presence of non-zero
terms on the left and total time derivatives in the right side of (\ref{dhq1}).

Consider a special case of the partial Hamiltonian formalism, when these terms
(with second derivatives) vanish, and call it nondynamical in the noncanonical
sector. This requires the following conditions on the Hamiltonians
\begin{equation}
\dfrac{\partial H_{0}}{\partial\dot{q}^{\beta}}=0,\ \ \ \ \dfrac{\partial
H_{\alpha}}{\partial\dot{q}^{\beta}}=0,\ \ \ \ \alpha,\beta=n_{p}+1,\ldots,n.
\label{hh}%
\end{equation}
Then (\ref{dhq1}) will have only the right side, which can be written as
\begin{equation}
\left(  \dfrac{\partial H_{\beta}}{\partial q^{\alpha}}-\dfrac{\partial
H_{\alpha}}{\partial q^{\beta}}+\left\{  H_{\beta},H_{\alpha}\right\}
\right)  \dot{q}^{\beta}=-\left(  \dfrac{\partial H_{0}}{\partial q^{\alpha}%
}-\dfrac{\partial H_{\alpha}}{\partial t}+\left\{  H_{0},H_{\alpha}\right\}
\right)  , \label{h1}%
\end{equation}
which is a system of linear algebraic equations for the noncanonical
velocities $\dot{q}^{\alpha}$ for given Hamiltonians $H_{0}$, $H_{\alpha}$. As
in (\ref{h1}) there are no noncanonical accelerations $\ddot{q}^{\alpha}$, so
on $TQ_{n-n_{p}}$ there is no real dynamics, if (\ref{hh}) is satisfied.

\section{Singular theories}

Let us consider in more detail the conditions (\ref{hh}) and express them in
terms of the Lagrangian. Using (\ref{hl0}) and the definition of the
additional Hamiltonians (\ref{ha}), we obtain
\begin{equation}
\dfrac{\partial^{2}L}{\partial\dot{q}^{\alpha}\partial\dot{q}^{\beta}%
}=0,\ \ \ \ \alpha,\beta=n_{p}+1,\ldots,n. \label{dl}%
\end{equation}
This means that the dynamics is described by a degenerate Lagrangian
(singular) theory, so that the rank $r_{W}$ of the Hessian matrix
\begin{equation}
W_{AB}=\left\Vert \dfrac{\partial^{2}L}{\partial\dot{q}^{A}\partial\dot{q}%
^{B}}\right\Vert ,\ \ \ \ A,B=1,\ldots,n \label{w}%
\end{equation}
is not only less than the dimension $n$ of the configuration space, but less
than or equal to the number of momenta (due to (\ref{dl}))
\begin{equation}
r_{W}\leq n_{p}. \label{rn}%
\end{equation}
In considering the strict inequality in (\ref{rn}) we find that the definition
of \textquotedblleft extra \textquotedblright\ $\left(  n_{p}-r_{W}\right)  $
momenta results in a $\left(  n_{p}-r_{W}\right)  $ relations, just as in the
Dirac theory of constraints \cite{dirac}, where there are $\left(
n-r_{W}\right)  $ (primary) constraints, if the standard Hamiltonian formalism
is used. It is important that the dimension $n$ of the configuration space and
the rank of the Hessian $r_{W}$ are fixed by the problem statement initially,
which does not allow a change to the number of constraints.

In the case of the partial Hamiltonian formalism the number $n_{p}$ is a free
parameter that can be chosen so that the constraints do not appear at all. To
do this, it is natural to equate the number of momenta with the rank of the
Hessian
\begin{equation}
n_{p}=r_{W}. \label{nr1}%
\end{equation}
As a result, the singular dynamics (theory with degenerate Lagrangians) can be
formulated in such a way that constraints will not occur (primary, secondary
or of higher level) \cite{dup2009, dup2011}. To do this, first, rename the
index of the Hessian matrix $W_{AB}$ (\ref{w}) so that the non-singular minor
rank $r_{W}$ is in the upper left hand corner, denote with the Latin alphabet
$i,j$ the first $r_{W}$ indices and with Greek letters $\alpha,\beta$ the
remaining $\left(  n-r_{W}\right)  $ indices. Next, we write the equations of
motion (\ref{dq1})--(\ref{dp1}), (\ref{h1}) as
\begin{align}
&  \dot{q}^{i}=\left\{  q^{i},H_{0}\right\}  +\left\{  q^{i},H_{\beta
}\right\}  \dot{q}^{\beta},\label{dq3}\\
&  \dot{p}_{i}=\left\{  p_{i},H_{0}\right\}  +\left\{  p_{i},H_{\beta
}\right\}  \dot{q}^{\beta},\label{dp3}\\
&  F_{\alpha\beta}\dot{q}^{\beta}=G_{\alpha}, \label{fq}%
\end{align}
where the index values are connected with the rank of the Hessian by
$i=1,\ldots,r_{W}$, $\alpha,\beta=r_{W}+1,\ldots,n$, and
\begin{align}
F_{\alpha\beta}  &  =\dfrac{\partial H_{\alpha}}{\partial q^{\beta}}%
-\dfrac{\partial H_{\beta}}{\partial q^{\alpha}}+\left\{  H_{\alpha},H_{\beta
}\right\}  ,\label{f}\\
G_{\alpha}  &  =D_{\alpha}H_{0}=\dfrac{\partial H_{0}}{\partial q^{\alpha}%
}-\dfrac{\partial H_{\alpha}}{\partial t}+\left\{  H_{0},H_{\alpha}\right\}  .
\label{dh}%
\end{align}
Note that the system of equations (\ref{dq3} )--(\ref{dh}) coincides with the
equations derived in the approach to singular theories, which uses the mixed
solutions of Clairaut's equation \cite{dup2009, dup2011} (except for the term
with the partial time derivative of $H_{\alpha}$ in (\ref{dh})). Equations
(\ref{dq3})--(\ref{fq}) are a system of first order differential equations for
the canonical coordinates $q^{i},p_{i}$, while with respect to the
noncanonical velocities $\dot{q}^{\alpha}$, this is an algebraic system.
Indeed, (\ref{fq}) is the usual system of linear equations with respect to
$\dot{q}^{\alpha}$, and the properties of its solutions can be used to
classify classical singular theories. We will consider only those cases, when
the system (\ref{fq}) is consistent, then there are two possibilities
determined by the rank of the matrix $F_{\alpha\beta}$:

\begin{enumerate}
\item Nongauge theory, when $\operatorname*{rank}F_{\alpha\beta}=r_{F}%
=n-r_{W}$ is full, so that the matrix $F_{\alpha\beta}$ is invertible. Then
from (\ref{fq}) we can determine all noncanonical velocities by
\begin{equation}
\dot{q}^{\alpha}=\bar{F}^{\alpha\beta}G_{\beta}, \label{q}%
\end{equation}
where $\bar{F}^{\alpha\beta}$ is the inverse matrix of $F_{\alpha\beta}$,
defined by the equation $\bar{F}^{\alpha\beta}F_{\beta\gamma}=F_{\gamma\beta
}\bar{F}^{\beta\alpha}=\delta_{\gamma}^{\alpha}$.

\item Gauge theory, the rank of the $F_{\alpha\beta}$ is incomplete, that is,
$r_{F}<n-r_{W}$, and the matrix $F_{\alpha\beta}$ is noninvertible. In this
case, we can find from (\ref{fq}) only $r_{F}$ noncanonical velocities, while
the rest $\left(  n-r_{W}-r_{F}\right)  $ of the velocities remain arbitrary
gauge parameters that correspond to the symmetries of a singular dynamical
system. In the particular case $r_{F}=0$ (or zero matrix $F_{\alpha\beta}$)
from (\ref{fq}) we obtain
\begin{equation}
G_{\alpha}=0,
\end{equation}
and all the noncanonical velocities correspond to $\left(  n-r_{W}\right)  $
gauge parameters of the theory.
\end{enumerate}

In the first case (nongauge theory) all noncanonical velocities can be
excluded by (\ref{q}) and substituting them in (\ref{dq3})--(\ref{dp3}). Then
we obtain the Hamilton-like equations for nongauge singular systems
\begin{align}
\dot{q}^{i}  &  =\left\{  q^{i},H_{0}\right\}  _{nongauge},\label{qnon}\\
\dot{p}_{i}  &  =\left\{  p_{i},H_{0}\right\}  _{nongauge}, \label{pnon}%
\end{align}
where we have introduced a new (nongauge) bracket for the two dynamical
quantities $A,B$
\begin{equation}
\left\{  A,B\right\}  _{nongauge}=\left\{  A,B\right\}  +D_{\alpha}A\cdot
\bar{F}^{\alpha\beta}\cdot D_{\beta}B, \label{nong}%
\end{equation}
and $D_{\alpha}$ is defined in (\ref{dh}). From (\ref{qnon})--(\ref{pnon}), it
follows that the new nongauge bracket (\ref{nong}) uniquely determines the
evolution of any dynamical quantity $A$ in time
\begin{equation}
\dfrac{dA}{dt}=\dfrac{\partial A}{\partial t}+\left\{  A,H_{0}\right\}
_{nongauge}. \label{da}%
\end{equation}
It is important too that the nongauge bracket (\ref{nong}) has all the
properties of the Poisson bracket: it is antisymmetric and satisfies the
Jacobi identity. Therefore, the definition (\ref{nong}) can be considered as
some deformation of the Poisson bracket, but not for all $2n$ variables as in
the standard case, but only for $2r_{W}$ canonical coordinates $\left(
q^{i},p_{i}\right)  $, $i=1,\ldots,r_{W}$. It follows from (\ref{nong}) and
(\ref{da}) that, as in the standard case, if $H_{0}$ does not depend on time
explicitly, then it is conserved.

In the second case (gauge theory), only some of the noncanonical velocities
$\dot{q}^{\alpha}$ can be eliminated, the number of which is equal to the rank
$r_{F}$ of the matrix $F_{\alpha\beta}$, and the rest $\left(  n-r_{W}%
-r_{F}\right)  $ of the velocities are still arbitrary and can serve as gauge
parameters. Indeed, if the matrix $F_{\alpha\beta}$ is singular and of the
rank $r_{F}$, then we can bring it to the form that a non-singular minor of
size $r_{F}\times r_{F}$ which will be in the upper left-hand corner. Then in
the system (\ref{fq}), only the first $r_{F}$ equations are independent. Let
us present (\textquotedblleft noncanonical\textquotedblright) indices
$\alpha,\beta=r_{W}+1,\ldots,n$ in the form of pairs $\left(  \alpha
_{1},\alpha_{2}\right)  $, $\left(  \beta_{1},\beta_{2}\right)  $, where
$\alpha_{1},\beta_{1}=r_{W}+1,\ldots,r_{F}$ number the first $r_{F}$
independent rows of the matrix $F_{\alpha\beta}$ and correspond to the
nonsingular minor of $F_{\alpha_{1}\beta_{1}}$, the remaining $\left(
n-r_{W}-r_{F}\right)  $ rows will be dependent on the first ones, and
$\alpha_{2},\beta_{2}=r_{F}+1,\ldots,n$. Then the system (\ref{fq}) can be
written as
\begin{align}
F_{\alpha_{1}\beta_{1}}\dot{q}^{\beta_{1}}+F_{\alpha_{1}\beta_{2}}\dot
{q}^{\beta_{2}}  &  =G_{\alpha_{1}},\label{fg1}\\
F_{\alpha_{2}\beta_{1}}\dot{q}^{\beta_{1}}+F_{\alpha_{2}\beta_{2}}\dot
{q}^{\beta_{2}}  &  =G_{\alpha_{2}}. \label{fg2}%
\end{align}
Since $F_{\alpha_{1}\beta_{1}}$ is non-singular by construction, we can
express the first $r_{F}$ canonical velocities $\dot{q}^{\alpha_{1}}$ in terms
of the remaining $\left(  n-r_{W}-r_{F}\right)  $ velocities $\dot{q}%
^{\alpha_{2}}$ as follows%
\begin{equation}
\dot{q}^{\alpha_{1}}=\bar{F}^{\alpha_{1}\beta_{1}}G_{\beta_{1}}-\bar
{F}^{\alpha_{1}\beta_{1}}F_{\beta_{1}\alpha_{2}}\dot{q}^{\alpha_{2}},
\label{q1}%
\end{equation}
where $\bar{F}^{\alpha_{1}\beta_{1}}$ is $r_{F}\times r_{F}$-matrix which is
inverse to $F_{\alpha_{1}\beta_{1}}$. Further, due to the fact that
$\operatorname*{rank}F_{\alpha_{1}\beta_{1}}=r_{F}$, the other blocks can be
expressed via a non-singular block $F_{\alpha_{1}\beta_{1}}$%
\begin{align}
F_{\alpha_{2}\beta_{1}}  &  =\lambda_{\alpha_{2}}^{\alpha_{1}}F_{\alpha
_{1}\beta_{1}},\label{f1}\\
F_{\alpha_{2}\beta_{2}}  &  =\lambda_{\alpha_{2}}^{\alpha_{1}}F_{\alpha
_{1}\beta_{2}}=\lambda_{\alpha_{2}}^{\alpha_{1}}\lambda_{\beta_{2}}%
^{\gamma_{1}}F_{\alpha_{1}\gamma_{1}},\label{f2}\\
G_{\alpha_{2}}  &  =\lambda_{\alpha_{2}}^{\alpha_{1}}G_{\alpha_{1}}, \label{g}%
\end{align}
where $\lambda_{\alpha_{2}}^{\alpha_{1}}=\lambda_{\alpha_{2}}^{\alpha_{1}%
}\left(  q^{i},p_{i},q^{\alpha}\right)  $ are $r_{F}\times\left(
n-r_{W}-r_{F}\right)  $ smooth functions. Since the matrix $F_{\alpha\beta}$
is given, we can determine the function $\lambda_{\alpha_{2}}^{\alpha_{1}}$ by
$r_{F}\times\left(  n-r_{W}-r_{F}\right)  $ equations (\ref{f1})
\begin{equation}
\lambda_{\alpha_{2}}^{\alpha_{1}}=F_{\alpha_{2}\beta_{1}}\bar{F}^{\alpha
_{1}\beta_{1}}.
\end{equation}
Because $\left(  n-r_{W}-r_{F}\right)  $ velocities $\dot{q}^{\alpha_{2}}$ are
arbitrary, we can put them equal to zero
\begin{equation}
\dot{q}^{\alpha_{2}}=0,\ \ \ \ \alpha_{2}=r_{F}+1,\ldots,n, \label{q2}%
\end{equation}
which can be considered as a gauge condition. Then from (\ref{q1}), it follows
that
\begin{equation}
\dot{q}^{\alpha_{1}}=\bar{F}^{\alpha_{1}\beta_{1}}G_{\beta_{1}},\ \ \ \ \alpha
_{1}=r_{W}+1,\ldots,r_{F}.
\end{equation}
By analogy with (\ref{nong}), we introduce a new (gauge) bracket
\begin{equation}
\left\{  A,B\right\}  _{gauge}=\left\{  A,B\right\}  +D_{\alpha_{1}}A\cdot
\bar{F}^{\alpha_{1}\beta_{1}}\cdot D_{\beta_{1}}B. \label{gaug}%
\end{equation}
Then the equations of motion (\ref{dq3})--(\ref{fq}) can be written in the
Hamiltonian-like form (as (\ref{qnon} )--(\ref{pnon} ))
\begin{align}
\dot{q}^{i}  &  =\left\{  q^{i},H_{0}\right\}  _{gauge},\\
\dot{p}_{i}  &  =\left\{  p_{i},H_{0}\right\}  _{gauge}.
\end{align}
The evolution of a physical quantity $A$ in time, as (\ref{da}), is determined
by the gauge bracket (\ref{gaug})%
\begin{equation}
\dfrac{dA}{dt}=\dfrac{\partial A}{\partial t}+\left\{  A,H_{0}\right\}
_{gauge}.
\end{equation}
In the particular case, when $r_{F}=0$, we have
\begin{equation}
F_{\alpha\beta}=0, \label{f0}%
\end{equation}
and hence all additional Hamiltonians vanish $H_{\alpha}=0$, then it is seen
from the definition (\ref{ha}) that the Lagrangian does not depend on the
noncanonical velocities $\dot{q}^{\alpha}$. Therefore, taking into account
(\ref{f0}) and (\ref{fq}) we find that the partial Hamiltonian $H_{0}$ does
not depend on the noncanonical generalized coordinates $q^{\alpha}$
\begin{equation}
\dfrac{\partial H_{0}}{\partial q^{\alpha}}=0,
\end{equation}
if $H_{0}$ is manifestly independent of time. In this case, the gauge bracket
coincides with the Poisson bracket, since the second term in (\ref{gaug})
vanishes. Thus, we have shown that the singular theories (with degenerate
Lagrangians) at the classical level can be described in terms of the partial
Hamiltonian formalism with the number of momenta of $n_{p}$, equal to the rank
$r_{W}$ of the Hessian matrix $n_{p}=r_{W}$, without the introduction of
additional relationships between the dynamical variables (constraints).

\section{Origin of constraints in singular theories}

As noted above (after (\ref{rn})), the introduction of additional dynamical
variables must necessarily give rise to additional relationships between them.
For example, let us consider the \textquotedblleft extra\textquotedblright%
\ $\left(  n-r_{W}\right)  $ momenta $p_{\alpha}$ (since we have a complete
description of the dynamics without them), which correspond to noncanonical
generalized velocities $\dot{q}^{\alpha}$ for the standard definition
\cite{dirac}
\begin{equation}
p_{\alpha}=\dfrac{\partial L}{\partial\dot{q}^{\alpha}},\ \ \ \ \ \alpha
=r_{W}+1,\ldots,n. \label{pa}%
\end{equation}
So (\ref{pa}), together with the definition of the partial generalized
canonical momenta (\ref{pp}) coincide with the standard definition of the
\textquotedblleft full\textquotedblright\ momenta (\ref{p}). Using the
definition of additional Hamiltonians (\ref{ha}), we get the same $\left(
n-r_{W}\right)  $ relations
\begin{equation}
\Phi_{\alpha}=p_{\alpha}+H_{\alpha}=0,\ \ \ \ \alpha=r_{W}+1,\ldots,n,
\label{fa}%
\end{equation}
which are called the (primary) constraints \cite{dirac} (in a resolved form).
These relations (\ref{fa}) are similar to the standard procedure of extension
of the phase space (\ref{pn1}). One can enter any number $n_{p}^{\left(
add\right)  }$ of \textquotedblleft extra\textquotedblright\ momenta $0\leq
n_{p}^{\left(  add\right)  }\leq n-r_{W}$, then the theory will have the same
number $n_{p}^{\left(  add\right)  }$ of (primary) constraints. In the partial
Hamiltonian formalism we have considered the case of $n_{p}^{\left(
add\right)  }=0$, while in the Dirac theory, $n_{p}^{\left(  add\right)
}=n-r_{W}$, although it is possible to take intermediate variants, to solve a
specific task.

Now, the transition to the Hamiltonian by the standard formula
\begin{equation}
H_{total}=p_{i}\dot{q}^{i}+p_{\alpha}\dot{q}^{\alpha}-L, \label{ht}%
\end{equation}
cannot be done directly\footnote{Therefore, $H_{total}$ is also not a
\textquotedblleft true\textquotedblright\ Hamiltonian, as $H_{0}$ in
(\ref{hl0}).}, because it is impossible to express the noncanonical velocities
$\dot{q}^{\alpha}$ through the \textquotedblleft extra\textquotedblright%
\ momenta $p_{\alpha}$ and then apply the standard Legendre transformation.
But it is possible to transform $H_{total}$ (\ref{ht}) in such a way that one
can use the method of undetermined coefficients \cite{dirac}. It is important
that the constraints $\Phi_{\alpha}$ do not depend on the noncanonical
generalized velocities $\dot{q}^{\alpha}$, as well as the Hamiltonians
$H_{0},H_{\alpha}$, because the rank of the Hessian matrix is equal to
$r_{W}$. So the total Hamiltonian can be written as%
\begin{equation}
H_{total}=H_{0}+\dot{q}^{\alpha}\Phi_{\alpha}, \label{ht1}%
\end{equation}
where $\dot{q}^{\alpha}$ play the role of undetermined coefficients. The
equations of motion can be written in a Hamiltonian-like form in terms of the
total Hamiltonian and the full Poisson bracket (\ref{abf}) as follows%
\begin{align}
dq^{A}  &  =\left\{  q^{A},H_{total}\right\}  _{full}\ dt,\label{qpf}\\
dp_{A}  &  =\left\{  p_{A},H_{total}\right\}  _{full}\ dt \label{qpf1}%
\end{align}
with the $\left(  n-r_{W}\right)  $ of additional conditions (\ref{fa}).
However, the equations (\ref{qpf})--(\ref{qpf1}) and (\ref{fa}) are not
sufficient to solve the problem: to find the equations for the undetermined
coefficients $\dot{q}^{\alpha}$ in (\ref{ht1}). These equations can be derived
from some additional principles, such as conservation relations (\ref{fa}) in
time \cite{dirac}
\begin{equation}
\dfrac{d\Phi_{\alpha}}{dt}=0.
\end{equation}
The time dependence of any physical quantity $A$ is now determined by the
total Hamiltonian and the full Poisson bracket
\begin{equation}
\dfrac{dA}{dt}=\dfrac{\partial A}{\partial t}+\left\{  A,H_{total}\right\}
_{full}. \label{da1}%
\end{equation}
If the constraints do not depend explicitly on time, then from (\ref{da1}) and
(\ref{ht1}) we obtain
\begin{equation}
\left\{  \Phi_{\alpha},H_{total}\right\}  _{full}=\left\{  \Phi_{\alpha}%
,H_{0}\right\}  _{full}+\left\{  \Phi_{\alpha},\Phi_{\beta}\right\}
_{full}{\dot{q}}^{\beta}=0, \label{fh}%
\end{equation}
which is a system of equations for the undetermined coefficients $\dot
{q}^{\alpha}$. Note that (\ref{fh}) coincides with (\ref{fq}), because
\begin{align}
F_{\alpha\beta}  &  =\left\{  \Phi_{\alpha},\Phi_{\beta}\right\}
_{full},\label{fdh}\\
D_{\alpha}H_{0}  &  =\left\{  \Phi_{\alpha},H_{0}\right\}  _{full}.
\label{fdh1}%
\end{align}
However, in contrast to the reduced description (without the $\left(
n-r_{W}\right)  $ \textquotedblleft extra\textquotedblright\ momenta
$p_{\alpha}$), where (\ref{fq}) is a system of $\left(  n-r_{W}\right)  $
linear equations in $\left(  n-r_{W}\right)  $ unknowns $\dot{q}^{\alpha}$,
the extended system (\ref{fh}) can lead to additional constraints (of higher
stages), which significantly complicates the analysis of the physical dynamics
\cite{dirac,sundermeyer}.

It follows from (\ref{fdh})--(\ref{fdh1}), that the new brackets (gauge
(\ref{gaug}) and nongauge (\ref{nong})) transform into the corresponding Dirac
brackets. We also note that our classification on the gauge and nongauge
theories corresponds to the first- and second-class constraints \cite{dirac},
and the limiting case of $F_{\alpha\beta}=0$ (\ref{f0}) corresponds to Abelian
constraints \cite{gog/khv/per,lor05}.

\section{Conclusion}

Thus, in this paper a \textquotedblleft shortened\textquotedblright%
\ formulation of classical singular theories is given. In it there is no
concept of constraints, because \textquotedblleft extra\textquotedblright%
\ dynamical variables, namely, the generalized momenta corresponding to
noncanonical coordinates, are not introduced. To this purpose, a partial
Hamiltonian formalism is proposed. It is shown that a special case of it
effectively describes multi-time dynamics. It is proved that singular theories
(with degenerate Lagrangians) can be described in the framework of our
approach without the introduction of additional relations between the
dynamical variables (constraints), if the number of canonical generalized
momenta coincides with the rank of the Hessian matrix $n_{p}=r_{W}$. From a
physical point of view, the introduction of the \textquotedblleft
extra\textquotedblright\ momenta is not necessary, because there is no
dynamics in these (degenerate) directions.

The Hamiltonian formulation of singular theories is done by introducing new
brackets (gauge (\ref{gaug}) and nongauge (\ref{nong})), which have all the
properties of the Poisson brackets (antisymmetry, Jacobi identity and their
appearance in the equations of motion and evolution of the system in time).

If one extends the phase space to the full phase space, these brackets become
the Dirac brackets, and constraints are imposed on the \textquotedblleft
extra\textquotedblright\ momenta.

Our analysis suggests that the quantization of singular systems under the
proposed \textquotedblleft shortened\textquotedblright\ approach can be
carried out in a standard way, while not all $2n$ variables of the extended
phase space will be quantized, but only $2r_{W}$ variables of the (initially)
reduced phase space. The remaining (noncanonical) variables can be treated as
continuous parameters.

\bigskip

\section*{Acknowledgments}

The author is grateful to I. H. Brevik, G.~A.~Goldin, U.~G\"{u}nther,
M.~Hewitt, I.~Kanatchikov, G. Ch. Kurinnoy, A.~Nurmagambetov,
A.~A.~Voronov and M. Walker for useful discussions. He would like to express deep
thankfulness to Jim Stasheff for carefully reading all versions and making many
corrections and important remarks, and to S.~Deser for pointing out
\cite{arn/des/mis} reprinted as arXiv:gr-qc/0405109. Also, the author 
is thankful to the Alexander von Humboldt
Foundation for funding his research stay at the University of M\"unster,
and J. Cuntz for kind hospitality. 

\appendix\setcounter{section}{0} 
\renewcommand{\thesection}{\Alph{section}}
\def\theequation{\Alph{section}.\arabic{equation}}

\section{Appendix: Singular theory as multi-time dynamics}

The condition (\ref{hh}) means that the Hamiltonians do not depend explicitly
on the noncanonical velocities $\dot{q}^{\alpha}$, that is, $H_{0}%
=H_{0}\left(  t,q^{i},p_{i},q^{\alpha}\right)  $, $H_{\alpha}=H_{\alpha
}\left(  t,q^{i},p_{i},q^{\alpha}\right)  $, $\alpha=n_{p}+1,\ldots,n$. Thus,
the dynamical problem is defined on the manifold $TQ_{n_{p}}^{\ast}\times
Q_{n-n_{p}}$, so that $q^{\alpha}$ actually play the role of real parameters,
similar to the time\footnote{In this case the nondynamical $Q_{n-n_{p}}$ is
isomorphic to the real space $R^{n-n_{p}}$.}.

Recalling (\ref{qn1}) and reversing it, we can interpret $\left(
n-n_{p}\right)  $ canonical generalized coordinates $q^{\alpha}$ as $\left(
n-n_{p}\right)  $ \textquotedblleft extra\textquotedblright\ (to $t$) times,
and $H_{\alpha}$ as $\left(  n-n_{p}\right)  $ corresponding Hamiltonians.
Indeed, we introduce the notation
\begin{align}
\tau^{\mu}  &  =t,\ \ \ \ \mathsf{H}_{\mu}=H_{0},\ \ \ \ \ \mu=0,\\
\tau^{\mu}  &  =q^{\mu+n_{p}},\ \ \ \ \mathsf{H}_{\mu}=H_{\mu+n_{p}%
},\ \ \ \ \mu=1,\ldots,\left(  n-n_{p}\right)  ,
\end{align}
where $\mathsf{H}_{\mu}=\mathsf{H}_{\mu}\left(  \tau_{\mu},q^{i},p_{i}\right)
$ are Hamiltonians of the multi-time dynamics with $n_{\mu}=n-n_{p}+1$ times
$\tau^{\mu}$. Note that $\tau^{\mu}$ can be called generalized times, because
they are not real times (for the real multi-time physics see
\cite{dor70,dva/gab/sen} and review \cite{bar/ter}), in the same way as
generalized coordinates have nothing to do with space-time coordinates
\cite{lan/lif1}.

In \ this formulation, the differential of the action $\mathsf{S}%
=\mathsf{S}\left(  \tau^{\mu},q^{i}\right)  $ in the multi-time dynamics can
be written as \cite{lon/lus/pon} (see also \cite{arn/des/mis})%
\begin{equation}
d\mathsf{S}=p_{i}dq^{i}-\mathsf{H}_{\mu}d\tau^{\mu},\ \ i=1,\ldots
,n_{p},\ \ \mu=0,\ldots,\left(  n-n_{p}\right)  \ \label{dsp}%
\end{equation}
It follows that the partial derivatives of the action are
\begin{equation}
\dfrac{\partial\mathsf{S}}{\partial q^{i}}=p_{i},\ \ \ \dfrac{\partial
\mathsf{S}}{\partial\tau^{\mu}}=-\mathsf{H}_{\mu}, \label{sph}%
\end{equation}
and the system of $\left(  n-n_{p}+1\right)  $ Hamilton-Jacobi equations is
\begin{equation}
\dfrac{\partial\mathsf{S}}{\partial\tau^{\mu}}+\mathsf{H}_{\mu}\left(
\tau_{\mu},q^{i},\dfrac{\partial\mathsf{S}}{\partial q^{i}}\right)
=0,\ \ \ \ \mu=0,\ldots,\left(  n-n_{p}\right)  . \label{s1}%
\end{equation}
We note that from (\ref{sl}) and (\ref{sph}) it follows the relation between
$\mathsf{H}_{\mu}$. Indeed, we differentiate the Hamilton-Jacobi equations
(\ref{sl}) on $\tau^{\nu}$
\begin{align}
\dfrac{\partial^{2}\mathsf{S}}{\partial\tau^{\mu}\partial\tau^{\nu}}  &
=-\dfrac{\partial\mathsf{H}_{\mu}}{\partial\tau^{\nu}}-\dfrac{\partial
\mathsf{H}_{\mu}}{\partial p_{i}}\dfrac{\partial}{\partial\tau^{\nu}}%
\dfrac{\partial\mathsf{S}}{\partial q^{i}}=-\dfrac{\partial\mathsf{H}_{\mu}%
}{\partial\tau^{\nu}}-\dfrac{\partial\mathsf{H}_{\mu}}{\partial p_{i}}\left(
-\dfrac{\partial\mathsf{H}_{\nu}}{\partial q^{i}}-\dfrac{\partial
\mathsf{H}_{\nu}}{\partial p_{j}}\dfrac{\partial}{\partial q^{i}}%
\dfrac{\partial\mathsf{S}}{\partial q^{j}}\right) \nonumber\\
&  =-\dfrac{\partial\mathsf{H}_{\mu}}{\partial\tau^{\nu}}+\dfrac
{\partial\mathsf{H}_{\mu}}{\partial p_{i}}\dfrac{\partial\mathsf{H}_{\nu}%
}{\partial q^{i}}+\dfrac{\partial\mathsf{H}_{\mu}}{\partial p_{i}}%
\dfrac{\partial\mathsf{H}_{\nu}}{\partial p_{j}}\dfrac{\partial^{2}\mathsf{S}%
}{\partial q^{i}\partial q^{j}}. \label{d2s}%
\end{align}
Then antisymmetrization (\ref{d2s}) gives the integrability condition
\begin{equation}
\dfrac{\partial^{2}\mathsf{S}}{\partial\tau^{\nu}\partial\tau^{\mu}}%
-\dfrac{\partial^{2}\mathsf{S}}{\partial\tau^{\mu}\partial\tau^{\nu}}%
=\dfrac{\partial\mathsf{H}_{\mu}}{\partial\tau^{\nu}}-\dfrac{\partial
\mathsf{H}_{\nu}}{\partial\tau^{\mu}}+\left\{  \mathsf{H}_{\mu},\mathsf{H}%
_{\nu}\right\}  =0. \label{sth}%
\end{equation}
To obtain the equations of motion, one needs to set the variation to zero:
$\delta\mathsf{S}=0$, where%
\begin{equation}
\mathsf{S}=\int\left(  p_{i}dq^{i}-\mathsf{H}_{\mu}d\tau^{\mu}\right)  ,
\end{equation}
taking into account that independent variations $\delta q^{i}$, $\delta p_{i}%
$, $\delta\tau^{\mu}$ vanish at the ends of the interval of integration. We
obtain
\begin{equation}
\delta\mathsf{S}=\int\delta p_{i}\left(  dq^{i}-\dfrac{\partial\mathsf{H}%
_{\mu}}{\partial p_{i}}d\tau^{\mu}\right)  +\int\delta q^{i}\left(
-dp_{i}-\dfrac{\partial\mathsf{H}_{\mu}}{\partial q^{i}}d\tau^{\mu}\right)  ,
\end{equation}
where the equations of the Hamiltonian for the multi-time dynamics can be
written in differential form \cite{lon/lus/pon}
\begin{align}
dq^{i}  &  =\left\{  q^{i},\mathsf{H}_{\mu}\right\}  d\tau^{\mu},\label{dq2}\\
dp_{i}  &  =\left\{  p_{i},\mathsf{H}_{\mu}\right\}  d\tau^{\mu}, \label{dp2}%
\end{align}
which coincide with (\ref{dq})--(\ref{dp}) by construction. The integrability
conditions (\ref{sth}) can be also written in differential form\footnote{As in
the standard case, \cite{arnold}, the equations (\ref{dq2} )--(\ref{hh1}) are
the conditions for closure of a differential 1-form (\ref{dsp}).}
\begin{equation}
\left(  \dfrac{\partial\mathsf{H}_{\mu}}{\partial\tau^{\nu}}-\dfrac
{\partial\mathsf{H}_{\nu}}{\partial\tau^{\mu}}+\left\{  \mathsf{H}_{\mu
},\mathsf{H}_{\nu}\right\}  \right)  d\tau^{\nu}=0,\ \ \ \ \mu,\nu
=0,\ldots,\left(  n-n_{p}\right)  \label{hh1}%
\end{equation}
which coincide with the equations (\ref{h1}), also written in differential form.

Thus, we have shown that the nondynamical sector in the noncanonical version
of the partial Hamiltonian formalism (which is determined by the equations of
motion (\ref{dq1})--(\ref{dhq1}) with the additional integrability conditions
(\ref{hh}), but without any conditions on the number of moments $n_{p}$) can
be formulated as the multi-time dynamics with the number of generalized times
(and corresponding Hamiltonians $\mathsf{H}_{\mu}$) $n_{\mu}=n-n_{p}+1$ and
the equations (\ref{dq2})--(\ref{hh1}). In this formulation the number of
generalized times $n_{\mu}$ is not fixed, and $1\leq n_{\mu}\leq n+1$, because
the number of generalized momenta $n_{p}$ is arbitrary less than or equal to the
dimension of the configuration space $n$. They are connected by the relation%
\begin{equation}
n_{\mu}+n_{p}=n+1, \label{nn}%
\end{equation}
which can be called a times-momenta rule. In the particular case of singular
theories (with degenerate Lagrangians), the number of momenta $n_{p}$ is fixed
by the condition (\ref{nr1}) $n_{p}=r_{W}$. So from (\ref{nn}) we obtain%
\begin{equation}
n_{\mu}+r_{W}=n+1, \label{nr}%
\end{equation}
which can be called a times-rank rule. If (\ref{hh}) and (\ref{nr}) are
fulfilled, then such a singular theory can be (effectively) described by the
multi-time dynamics.


\end{document}